# Bublz! : Playing with Bubbles to Develop Mathematical Thinking

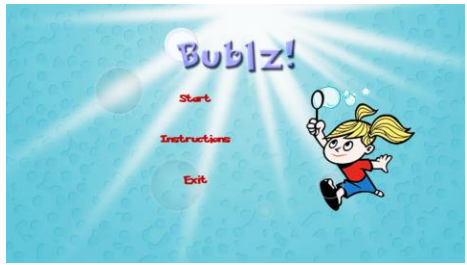

**Figure 1**: Bublz! Start Screen

**Dhruv Chand**
National Institute of Technology Karnataka
Surathkal, Karnataka, India
dhruvchand@live.com

**Karthik Gopalakrishnan**
Indian Institute of Technology Patna
Patna, Bihar, India
karthik.cs11@iitp.ac.in

**Nisha K K**
National Institute of Technology Karnataka
Surathkal, Karnataka, India
nishakk94@gmail.com

**Mudit Sinha**
Vellore Institute of Technology
VIT University
Chennai, Tamil Nadu, India
mudit.sinha2012@vit.ac.in

**Shreya Sriram**
Delhi Technological University
New Delhi, India
shreyasriram29@gmail.com

## Abstract
We encounter mathematical problems in various forms in our lives, thus making mathematical thinking an important human ability [4]. In this paper, we present Bublz!, a simple, click-driven game for children to engage in and develop mathematical thinking in an enjoyable manner.

## Author Keywords
Game Design; Mathematics; Reasoning; Mathematical Thinking; Strategic Thinking;

## ACM Classification Keywords
K.3.1;

## Introduction
A complete understanding of mathematics not only includes knowledge of mathematical concepts and principles, but also the capacity to engage in mathematical thinking: observing patterns, examining constraints, making conjectures and inferences, and so on [3]. Mathematics is a dynamic process of "gathering, discovering and creating knowledge in the course of some activity having a purpose" [2]. Such activities or tasks typically have more than one solution strategy and involve decision-making and interpreting the reasonableness of possible actions [5].

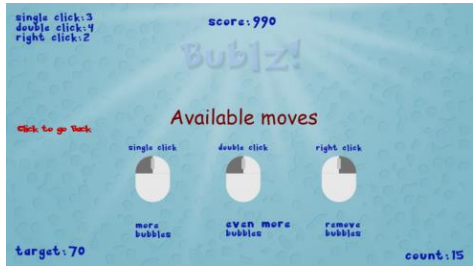

**Figure 2**: Instructions Screen

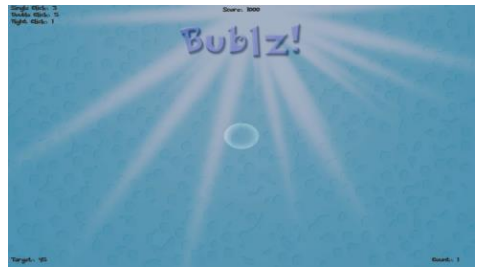

**Figure 3**: Main game screen at the start of a level, showing one bubble to start making moves on

Bublz! (Figure 1) is a single-player game in which players are given a task of this nature in each game level. The players need to do simple arithmetic and effectively reason with numbers, which are depicted using bubbles, in order to complete the task. The only mathematical prerequisite to play the game is the ability to do addition and subtraction, which are depicted through the creation and removal/popping of bubbles respectively. We developed a fully-functioning digital prototype for the task with six levels of play using Unity 3D [6] in the JavaScript and C# programming languages. The game is primarily intended to be played on a computer, but it can very easily be modified and built to work on portable devices like phones and tablets as well.

### Game Play
For each of the six levels in the game, a unique triplet of positive numbers ($L$, $D$, $R$) is associated with the three common types of clicks done on a mouse: single left-click, double left-click and single right-click (Figure 2).

When a level starts, one bubble is displayed on the screen (Figure 3). A positive target number $T$ randomly chosen from the integers between 2 and 70, both inclusive, and the triplet associated with the three types of clicks for the level, are also displayed on the screen. The player is given an initial score of 1000.

- When the player does a single left-click on a bubble, $L$ new bubbles are created.

- When the player does a double left-click on a bubble, $D$ new bubbles are created.

- When the player does a single right-click on a bubble, $R$ bubbles are removed or "popped", including the bubble on which the click was done.

The player's task is to obtain $T$ bubbles on the screen using a combination of the three types of clicks, which are the possible types of moves. To obtain the highest possible score of 1000, the player should do it in the least number of moves in which $T$ bubbles can be obtained on the screen. With each move made beyond the least number of moves, the player's score reduces by 10.

Feedback on performance (Figure 4) is provided to players upon completing a level to facilitate learning [1]: the number of moves in which the player obtained $T$ bubbles and the minimum number of moves in which the player could have obtained it are displayed on the screen. The player can retry the same target, repeat the level with a different target, or try targets in the next level. A video trailer of the game is available at the following link: vimeo.com/132009026

### Game Viewed as an Optimization Problem
The task for a player in each level is to minimize the total number of moves made in order to maximize the score – an optimization problem. Let ($L$, $D$, $R$) be the triplet for a given level. Let $T$ be the displayed target. Let the number of single left-clicks, double left-clicks and single right-clicks done to create $T$ bubbles be $x$, $y$ and $z$ respectively. In order to obtain the highest score in a given level, the total number of moves must be minimized, hence the formulation of the optimization problem is:

Minimize $N = x + y + z$, such that

$$1 + Lx + Dy - Rz = T$$

$x$, $y$, $z$ are non-negative integers

### Game Design

For each level, the triplet values (*L*, *D*, *R*) are chosen such that any positive number can be generated using a linear combination of *L*, *D* and –*R*. For example, (3, 4, 1) is a valid triplet because any positive number can be generated using a linear combination of 3, 4 and -1. This ensures that there is always at least one sequence of moves to generate *T* bubbles.

Intuition suggests more bubbles should be generated on doing a double left-click since the effort put in is twice the effort put in to do a single left-click. Hence, with the aim of making the game easy to grasp, *L* and *D* are chosen such that *D* > *L*.

Triplet values of higher magnitudes are chosen for higher levels to make them more difficult. For example, if the current level has (3, 4, 1) as the click triplet, the next level could have (5, 7, 4) as the click triplet.

To ensure that there will always be at least one bubble on the screen that can be clicked, the game prevents the player from doing a single right-click move when the number of bubbles on the screen is lesser than or equal to the value *R*. This is done by displaying an appropriate message (Figure 5). To ensure that the player doesn't unnecessarily overload the screen with bubbles by doing single/double left-clicks all the time, the game also prevents the player from making a single/double left-click move if making that move would result in the total number of bubbles on the screen exceeding a threshold value of 150. This additionally serves as a hint to the player that the target can be reached without ever exceeding 150 bubbles.

### Evaluation

Our immediate goal after developing the game was to get feedback from children and evaluate our design choices. Hence, we organized a preliminary play session for 10 middle school students from a local school in Bangalore, India, and observed while they played. The play session involved playing three successive levels of the game.

We made several observations from the play session. As the play session progressed, most students who got large targets almost instantly did a double left-click as their first move. This suggests that in the course of playing the game, the students developed a mental representation associating the number of left-clicks with the number of bubbles they could generate, a positive feedback to our design decision of generating more bubbles for a double left-click than for a single left-click. The students also generally took greater time to complete higher levels and additionally reported during post-play feedback interviews that they found higher levels challenging, a positive feedback to our decision to use triplets of higher magnitudes in higher levels. No student overloaded the screen with bubbles, making the upper cap of 150 on the number of bubbles seem unnecessary. We observed the time students spent playing the game and determined their intention to replay the game through post-play interviews. According to both factors, students showed clear motivation to play the game. They reported liking and enjoying the game. Some students wanted the game to be more visually appealing while others wanted more levels to play, which we took into account to improve the game. When asked about the game's utility to them, most students felt the game would make them "better at mathematics".

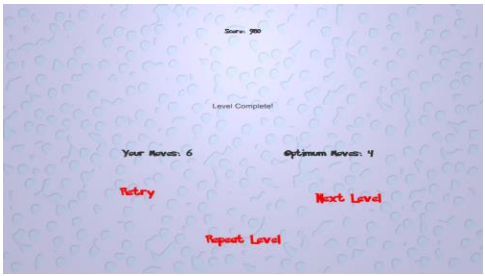

**Figure 4**: Score screen after finishing a level, showing the score, number of moves made, optimal number of moves, and various game options

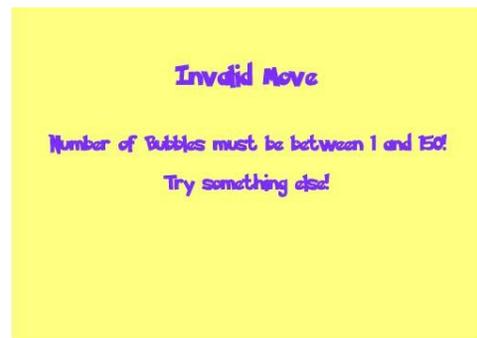

**Figure 5**: Invalid move screen is displayed when an attempted move would cause the number of bubbles to go outside the range 1-150

**Future Work**
We plan to make refinements to the game, such as using achievement stars instead of numeric scores, a popular mechanic in many games at present, and also removing the count of bubbles present on the screen to make it more challenging. We also plan to undertake a formal study of the impact that regular play of the game has on mathematical thinking in children. One way is to study the correlation that known measures of mathematical thinking have with an edit-distance like metric that tells how close a child's performance is at any point of time to that of an ideal player using in-game performance data. It would also be interesting to see whether regularly playing the game leads to improved performance on optimization problems encountered in later stages of life, such as active stock trading with the goal of maximizing profit.

**Acknowledgements**
We thank Amy Ogan, Erin Walker and Erik Harpstead for their guidance and assistance in developing Bublz!. We thank MS Ramaiah Institute of Technology, Bangalore, India, for enabling us to play-test our game with children. We also thank Carolyn P. Rosé and Carnegie Mellon's Internship Program in Technology Supported Education, which brought us together for 15 days to conceptualize, develop and conduct a preliminary evaluation of this game.

**References**
[1] Anderson, J. R., Corbett, A. T., Koedinger, K. R., & Pelletier, R. (1995). Cognitive Tutors: Lessons Learned. *The Journal of the Learning Sciences*, 4(2), 167-207.

[2] Romberg, T. A. (1992). Perspectives on Scholarship and Research Methods. *Handbook of research on mathematics teaching and learning* (pp. 49-64).

[3] Schoenfeld, A. H. (1992). Learning to Think Mathematically: Problem Solving, Metacognition and Sense Making in Mathematics. *Handbook of research on mathematics teaching and learning* (pp. 334-371).

[4] Stacey, K. (2006). What is Mathematical Thinking and Why is it Important? *Progress report of the APEC project: Collaborative Studies on Innovations for Teaching and Learning Mathematics in Different Cultures (II)–Lesson Study focusing on Mathematical Thinking*.

[5] Stein, M. K., Grover, B. W., Henningsen, M. (1996). Building Student Capacity for Mathematical Thinking and Reasoning: An Analysis of Mathematical Tasks used in Reform Classrooms. *American Educational Research Journal*.

[6] The Unity Game Engine, http://unity3d.com